\documentclass[12pt]{article}
\usepackage{amsthm,graphicx,amssymb}
\newcommand{\rmd}{\mathrm{d}}
\newcommand{\kartink}[2]{\begin{figure}[h,t,b]\begin{center}
\includegraphics[width=0.9\textwidth]{#1.eps}
\end{center}
\caption{#2}\end{figure}}
\newcommand{\ssyl}[6]{\bibitem{#1}#2, #3\ \textbf{#4}, #6 (#5).}

\author{S. V. Krasnikov\thanks{The Central Astronomical Observatory of RAS,
M-140, Pul\-ko\-vo, St.~Petersburg, Russia. \emph{Email}:
Gennady.Krasnikov@pobox.spbu.ru}}
\title{Gravitational strings. Do we see one?}
\date{}
\begin{document}
\maketitle
\begin{abstract}
I present a class of objects called gravitational strings (GS) for their
similarity to the conventional  cosmic strings:  even though  the former
are just singularities in flat spacetime, both varieties are equally
``realistic", they may play an equally important cosmological r\^ole and
their lensing properties are akin. I argue that the enigmatic object
CSL-1 is an evidence  in favor of the existence of GS.
\end{abstract}
\section*{Introduction}

The objects discussed below | the gravitational strings | have much in
common with the conventional  cosmic strings. So, I begin with a brief
reminder of a few key facts about the latter \cite{obz}.

The appearance of  the cosmic strings in the early universe is usually
explained as follows. Suppose, after the universe had cooled below some
critical temperature, a complex scalar field $\phi$ appeared with   the
Mexican-hat potential
\[
L=\partial_\mu\phi\partial^\mu\bar\phi - {\textstyle\frac14}(\phi\bar\phi
-a^2)^2,\qquad a> 0.
\]
One expects the field to take the minimal energy value, i.~e., $\phi=a
e^{i\sigma}$ with arbitrary $\sigma\in\mathbb R$. But the evolutions of
the field in different regions are
\emph{uncorrelated} and this can make the process
energetically prohibited even though \emph{locally} it is energetically
favourable. Indeed, if the field happened to develop a non-zero winding
number on some loop $C$, see figure~\ref{ris1},
\kartink{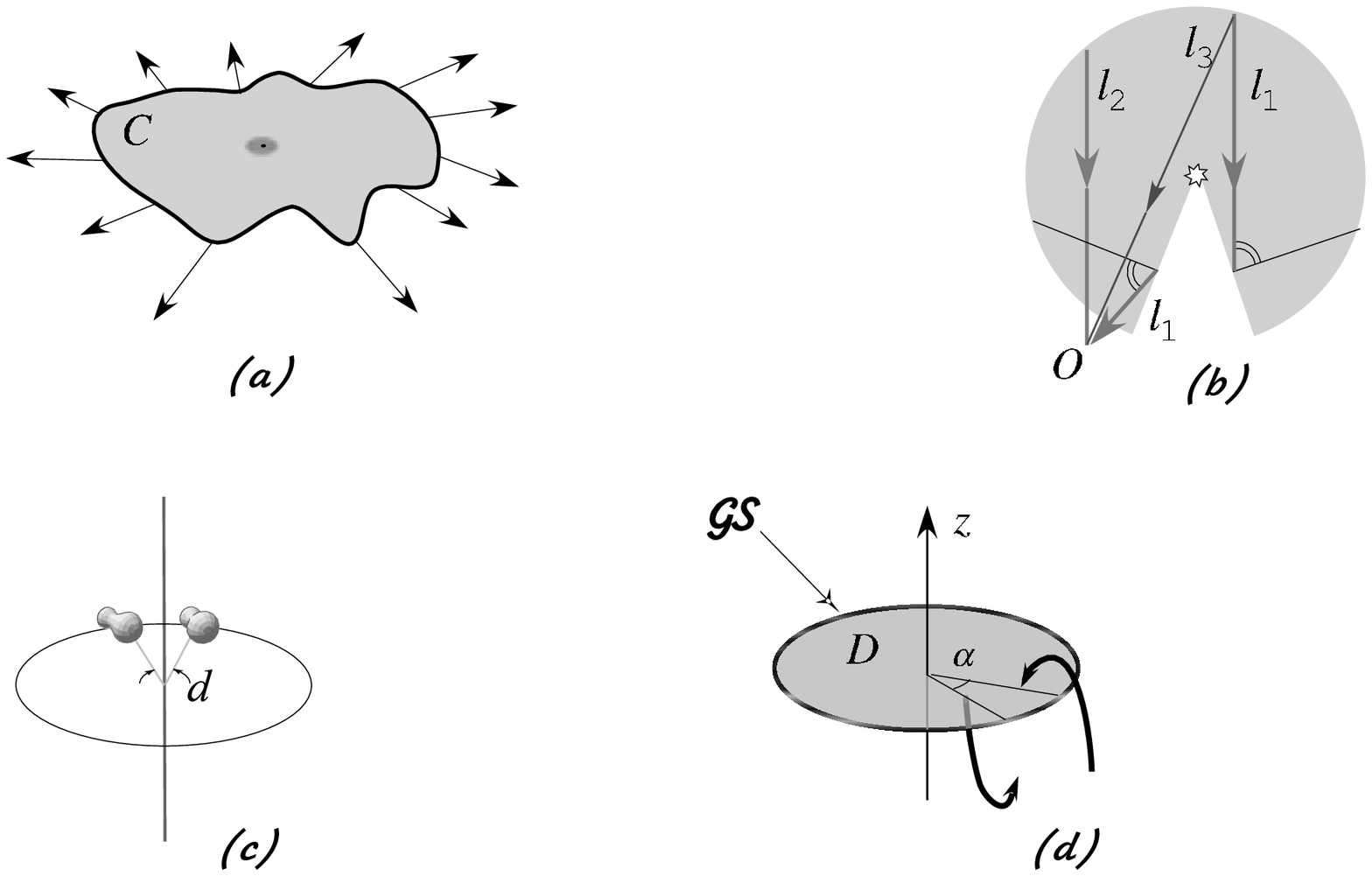}{\label{ris1} a) The arrows represent complex numbers, not
3-vectors. b) The ray $l_1$ is parallel to  $l_2$ and starts from the
same point as  $l_3$. Still, all three meet in $O$. c) The images of a
non-pointlike source are rotated w.~r.~t. each other. d) What looks as
two thick curves is, in fact, a single continuous curve.} then it cannot
be extended to a surface enclosed by $C$ without vanishing in some point.
Clearly, the energy density around that point will be non-zero. This fact
is purely topological, so if we continuously deform the loop or the
surface, the point with the non-zero energy density will persist. So, the
field configuration is an endless thin tube. The tube is stable: even
though it is surrounded by vacuum, it cannot dissolve for the topological
reasons discussed above. It is such tubes that are called (material)
cosmic strings.

The importance of cosmic strings, as well as their most promising (from
the observational point of view) manifestation stem from their | quite
unusual | gravitational fields. The universe  with a straight endless
string is believed to be described | \emph{at large $\rho$} | by the
spacetime
\begin{eqnarray}\label{1}
\rmd s^2= -\rmd t^2 + \rmd z^2 + \rmd \rho^2 + \rho^2\rmd \varphi^2,
\\
  t,z\in \mathbb R,\quad \rho>0,\quad \varphi=\varphi+2\pi-d,\quad d\in
  (0,2\pi).\nonumber
\end{eqnarray}
It is often convenient to represent the sections $t=const$ of this
spacetime as the results of the following surgery: a dihedral angle
having the $z$-axis as its edge is cut out of the Euclidean space
$\mathbb E^3$ and the half-planes bounding it are glued together (with no
shift along the axis). Since the axis cannot be glued back into the
spacetime, there is a singularity now ``in its place". This singularity
(often called conical) is sometimes associated in folklore with the
string itself.

When a string moves through the cosmological fluid it leaves a wake
behind it: as is seen from Fig.~\ref{ris1}b, two \emph{parallelly} moving
galaxies may nevertheless collide after a string has passed between them
| a phenomenon of obvious importance to cosmology. On the other hand, two
light rays emitted from the same source may, for exactly the same
reasons, come to an observer from different directions, see
Fig.~\ref{ris1}b. Thus, a string acts as a gravitational lens   producing
multiple images of a single object, see~Fig.~\ref{ris1}c.

\section{Gravitational strings}

Now let us  consider the purely gravitational case.

At the end of the Planck era the classical spacetime had emerged and
started to expand obeying the Einstein equations. By the time it could be
confidently called classical it was practically flat (by Planck
standards, anyway), so we can speak of emergence of a flat spacetime. One
can think, however, that remote regions evolved uncorrelatedly and the
locally fovourable process of becoming Minkowskian might be impeded by
some \emph{global} obstacles. For example, a circle lying in a newborn
(non-simply connected) flat region might happen to be too short (or too
long) for its radius of curvature. Exactly as with matter fields such
obstacles would give rise to singularities. This time, however, those
would be true \emph{geometric} singularities and of a rather unusual kind
at that: the spacetime is flat, so the singularities are not associated
with infinite curvature, or energy density, etc. It is such singularities
| more precisely   two-dimensional\footnote{The meaning of the term
``two-dimensional" is hopefully  clear intuitively. For a rigorous
definition see \cite{NCS}.} singularities in flat spacetime | that I call
gravitational strings (GS).

An example of a GS is the singularity in the spacetime~(\ref{1}) (the
latter must be understood this time literally, not as an approximation
valid at large $\rho$). Other examples can be obtained, if one of the
mentioned half-planes, before it is glued to the other, is shifted along
the $z$-axis
\cite{GL}, or the   $t$-axis
\cite{spin}, or, finally, is boosted in the $z$-direction \cite{Tod}. The
resulting spacetimes are quite different, but each contains an infinite
straight gravitational string at rest. Recently the list was supplemented
by more curious species \cite{NCS} including spiral strings,  rotating
straight strings, and even closed strings. The (sections $t=const$ of
the) latter are obtained | or, rather, are described | as follows. Remove
the closure of the circle
\[
 D:\quad x^2+y^2<1
 \]
from the Euclidean space $\mathbb E^3$, rotate one of the banks of the
thus obtained slit (either is  a copy of $D$) by some $\alpha\neq 2\pi$
w.~r.~t. to the $z$-axis, and, finally, glue the banks together. The
missing circumference  Bd$\,D$ (it cannot be glued back into the space)
is a GS, see~Fig.~\ref{ris1}d.

Thus, a new class of objects is introduced | the gravitational strings.
They have much in common with the conventional (matter) cosmic strings:
both species are equally ``realistic" (the reasons to believe in their
existence are much the same), equally important cosmologically (both
leave wakes) and have similar lensing properties. Still, in some respects
they are quite different. In particular, the relevant spacetime being
\emph{empty}, the evolution of a gravitational string, its form, its
gravitational field, etc.\ have nothing to do with properties of any
matter field, with the Nambu action, and so on.

\section{Observations}
Whether gravitational strings do exist is, ultimately, to be answered by
observations. So, what would one expect to observe if there is a circular
(this form is chosen, because it is easier to imagine the formation of a
\emph{finite} object) string of a galactic size located far enough from
the Earth and tilted by an angle~$\theta\sim 1$ (no fine tuning!)
relative to the line of sight?
\kartink{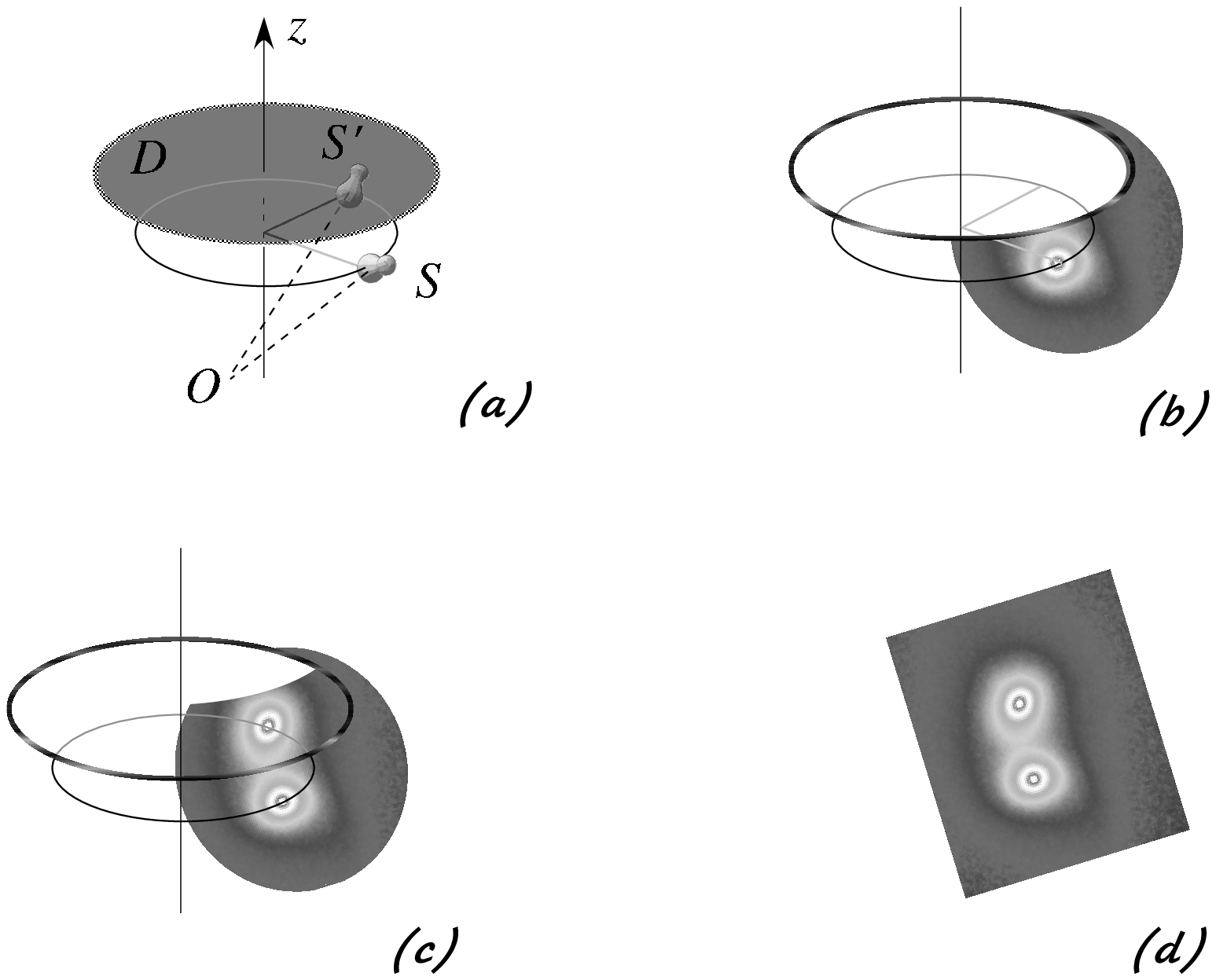}{a) $O$ sees two similar images. b) A single ``real"
galaxy. c) Its image in the presence of a loop GS. d) The Hubble Space
Telescope image of CSL-1.
\label{ris2} }

In analyzing this situation it is convenient to view the ``real" space as
the Euclidean space $\mathbb E^3$ (minus circumference) where light rays
are, as usual, straight lines, but where one additional rule holds: if a
ray meets $D$, its extension beyond $D$ is rotated by $\alpha$ w.~r.~t.
to the $z$-axis. So, to determine what the observer sees if there is a
bright source $S$ near  the string,  rotate $S$ by $\alpha$ obtaining
thus a ``fictituous" source $S'$, see Fig.~\ref{ris2}a.
 If one of the straight lines | $OS$ or $OS'$ | meets $D$ and the
other does not, then the observer will see both images $S$ and $S'$,
otherwise only one of them.
 Note
that the two images are not \emph{exactly} same: because $\theta\sim 1$,
they are rotated w.~r.~t. to each other. In summary: if there is an
appropriately located galaxy, see~Fig.~\ref{ris2}b, then what we must see
is two close galaxies with similar spectra, or in other words, two images
of the same galaxy as seen from different angles, see~Fig.~\ref{ris2}c.

Remarkably, such a picture \emph{is} observed \cite{CSL1}. A pair of
galaxies called CSL-1 are  1.9 arcsec separated and have the same spectra
identical at 98\% confidence level. This strongly suggests that what we
see (a HST image is shown in Fig.~\ref{ris2}d) is, in fact, two images of
the \emph{same} galaxy. The isophotes are not distorted, so the lensing
presumably does not involve a strong gravitational field, which rules out
a galaxy, or a conventional loop string as the lens. At the same time it
is not  a straight infinite string, because no more pairs have been found
in that region of the sky, while calculations predict at least 9 ones
\cite{ACSL}. It seems, thus, quite natural to interpret CSL-1 as an
observational proof of the existence of gravitational strings.

\end{document}